# A new heat source model for selective laser melting simulations based on energy distribution on the powder layer and the surface of substrate

Zhi Huang, Weibo Jia, Haoming Wang, Zhengtong Yang, Chao Li, Jie Liang, Yue Zhong.

School of Mechanical and Electrical Engineering, University of Electronic Science and Technology of China, Chengdu

**Corresponding author:**

Zhi Huang, University of Electronic Science and Technology of China, Chengdu 611731, China.

Email: zhihuang@uestc.edu.cn

**Abstract**

In order to predict the more accurate shape information of the melt pool in Selective Laser Melting (SLM), a new finite element temperature field simulations model is proposed. The simulations use a new heat source model that takes into account the influence of the powder layout, the surface of the substrate and the changes in the thickness of the powder layer after fusion on the energy distribution. In order to construct this new heat source model, firstly an improved optimization method based on the gradient descent and the univariate search technique is proposed to simulate the powder layout, and then the laser beam propagation between the powder and the surface of the substrate is tracked and recorded to obtain the energy distribution. Finally, according to the distribution of laser energy between the powder layer and the surface of the substrate, the heat source model is divided into two parts: one is the surface of substrate heat source model being the Gaussian distribution, the other one is the powder layer heat source model- satisfying the Gaussian distribution on the horizontal plane, changes in the depth direction according to the functional relationship obtained by the fitting. In addition, the thickness change of the powder layer after fusion is analyzed, and is taken into account in the heat source model. The powder simulation results are compared with the powder scattering experiment results to verify the effectiveness of the powder model. Comparing the temperature field simulation with the experiment, the results show that the predicted molten pool width relative error is 6.4%, and the connect width error is 9.6%, which has better accuracy and verifies the validity of the temperature field simulation model.

## 1.Introduction

With the continuous development and progress of additive manufacturing technology, selective laser melting technology (SLM) is increasingly used in many fields such as aerospace and medical treatment manufacturing. SLM technology has the advantages of short production cycle and high material utilization. The pros and cons of SLM parts are affected by the laser power, scanning speed, powder layer thickness, preheating temperature, scanning interval and other process parameters. The selection of process parameters has experiment and simulation prediction. Since the experiment is time-consuming and laborious, the low-cost simulation prediction is widely researched and applied.

The methods commonly used in the simulation of the temperature field of the SLM molten pool are the Finite Element Method (FEM) and the Volume of Fluid (VOF) method. In the finite element simulation process, the heat source model plays a vital role in the simulation results. Goldak [1] proposed a double ellipsoid heat source model in 1984. The model is composed of two quarter ellipsoids at the front and rear. In practical applications, it is necessary to observe the shape of the molten pool and adjust the parameters of double ellipsoid for getting a decent heat source model. However, the morphology of the molten pool is usually not easily and accurately captured. The cylindrical heat source model is also a widely applicable heat source model, which satisfies the Gaussian distribution on the cylindrical section, and gradually attenuates according to a certain law in the axial direction [2],[3]. The above models were used in the simulations, but the influence of powder layout on the heat source model was not considered.

In the VOF simulation process, the powder particles are usually regarded as spheres, and the arrangement of the powders has an important influence on energy distribution. At first, some researchers regarded the powder layer as uniformly sized spheres arranged in a simple cubic stacking method [4],[5]. Later, Galati proposed to use the spheres of various sizes to approximate the powders scattering [6]. Wu used the discrete element numerical method to simulate the generation of random powders [7], and validated the method according to the uniformity of statistical particle distribution. Tang used open source code LIGGGHTS to model the packing of powders in consideration of gravitational and contact forces [8]. Dai used the **direct algorithm compiled geometric model in MATLAB** software **to generate the powder bed** [9]. Khairallah used the ALE3D massively-parallel multi-physics code to target the set packing density [10]. The powder model is arranged randomly, without considering the influence of gravity, contact and other physical factors. The VOF method has high calculation accuracy, but due to its large amount of calculation, it is difficult to carry out large-scale simulation calculations [11].

Gusarov [12] combined the high efficiency of the FEM with the high precision of the VOF method, established a simulation model considering the absorption of laser energy in the assumed uniformly arranged powder. Tran

[13] considered the influence of particle radius and arrangement on energy absorption, proposed a new heat source model by counting the energy absorption rate of each layer applied to SLM FEM simulation. The heat source model has both a faster calculation speed and a higher accuracy. However, the effects of laser irradiation on the powder and on the surface of the substrate are not the same. The author did not distinguish between the two, and did not consider the influence of the change in the thickness of the powder layer after fusion on the simulations.

In summary, in order to make up for the shortcomings of the previous methods, this paper proposed a new heat source model that combines the advantages of the two simulation methods for SLM simulations which is more accurately predict the shape of the molten pool. The main research content of this paper is as follows: First, an improved optimization algorithm via the univariate search technique and the gradient descent method is proposed which is suitable for calculation the arrangement of the powder in the powder bed. Meanwhile this method avoids the errors caused by powder material parameter estimation. Second, experiments were used to test the effectiveness of the powder model. The obtained powder model was interacted with the laser to calculate and count the distribution of laser energy in the powder and the surface of the substrate. Considering the influence of the change of the powder layer thickness after fusion on the energy distribution, a heat source model that satisfies the Gaussian distribution in the horizontal direction and the vertical direction changes according to the function law is used to characterize the energy distribution in the powder, and the Gaussian heat source model is used to characterize the surface energy distribution of the substrate. Combine the two parts to form a new heat source model for FEM simulation. Finally, experiments were carried out to verify the accuracy of the prediction results.

## 2. Method proposed

*2.1. Optimization method to generate powder bed model*

Under the action of gravity and scraper squeezing, the powder layout has a tendency to move in the direction of the smallest potential energy. Therefore, this paper abstracts the powder layout problem as the optimization problem: The arrangement of the powder particles when the potential energy is the smallest. In order to solve the optimization problem, a new optimization algorithm is proposed by combined and improved the gradient descent and the univariate search technique. The powder layout model generated by this method does not need to consider the physical parameters, and avoids the error caused by the approximate parameters. Solving the optimization problem mainly includes four parts: the initial state of the powder, the objective function, the constraint conditions and the optimization algorithm.

*2.1.1. The initial state of the powder*

The initial state of the powder includes the quantity, the radius and the initial position. The specific methods are as follows:

The number of powder particles: The powder model produced in this research is only used for analysis and calculation of the laser energy distribution, so the powder model only needs to be filled within the range of laser irradiation, and the minimum number of powder particles $n_p$ is calculated by Eq. (1).

$$n_p = \frac{\pi r_0^2 h_{top}}{\frac{4}{3}\pi \bar{r}^3} \tag{1}$$

where $r_0$ is the beam waist of the laser pulse, $h_{top}$ is the thickness of the powder layer, $\bar{r}$ is the average radius of the powder.

Powder radius: The radius of the powder is determined by the particle size distribution.

The initial position of the powder particles: the powder particles is distributed on the horizontal plane centered at *(0, 0, $h_{top}$/2)*, and the interval between the centers of adjacent particles is *($h_{top}$/2+2s)*. In order to increase the degree of chaos of the initial particles, the particles are randomly displaced once in the horizontal plane, and the each powder displacement is *(random·s, random·s, 0)*.

Where *s* is the maximum distance of the particle along the coordinate axis, *random* is a random number between 0 and 1.

*2.1.2. Constraint conditions*

The powder particles are located above the substrate:

$$r_i - z_i \leq 0 \tag{2}$$

where $r_i$ is the radius of the i-th powder.

The powder particles are located below the flat surface scanned by the scraper.

$$z_i + r_i - h_{top} \leq 0 \tag{3}$$

Powder particles cannot be embedded in each other.

$$(x_i - x_j)^2 + (y_i - y_j)^2 + (z_i - z_j)^2 - (r_i + r_j)^2 \leq 0 \tag{4}$$

where *($x_i$, $y_i$, $z_i$)* are the coordinates of the i-th powder.

*2.1.3. Objective function*

The common tools for scattering powder are rollers or scrapers. The purpose of scattering powder is to keep the thickness of the powder layer consistent. In order to satisfy the requirements of layer thickness and small potential energy, this paper uses the sum of the square of the distance from the center of the powder to the center of the substrate as the objective function:

$$f(x, y, z) = \sum_{i=1}^{n}(x_i^2 + y_i^2 + z_i^2) \tag{5}$$

*2.1.4. Improved optimization algorithm*

Only using the univariate search technique or the gradient descent is easy to make the optimization problem enter the local optimum. This paper proposes a new algorithm that alternately used the univariate search technique and the gradient descent, and the step length is obtained by the maximum step length multiplied by a random number that could reduce the possibility of the algorithm entering the local optimum. The detailed optimization steps are as follows:

Step 1: Each particle moves in the horizontal plane along the steepest descent direction. Fig. 1 is a schematic diagram of powder movement. The direction of movement is determined by Eq. (6). The method for determining step length is shown in Fig. 2.

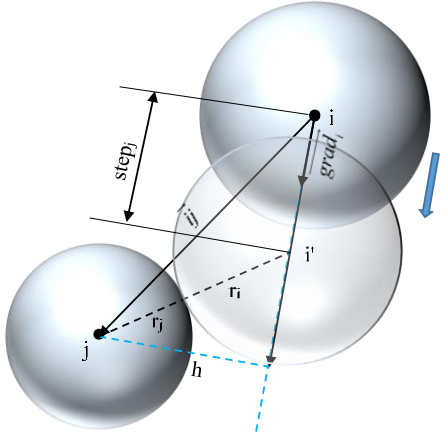

Fig. 1. Schematic diagram of powder movement.

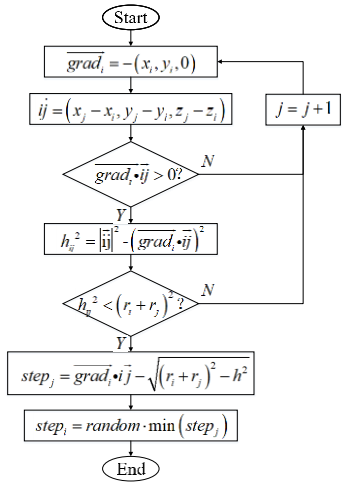

Fig. 2. Step length determination flow chart.

$$\overrightarrow{grad_i} = -(x_i, y_i, 0) \qquad (6)$$

where $\overrightarrow{grad_i}$ is the unit vector of the steepest descent direction of the i-th powder.

Step 2: Each particle moves in the direction in which the objective function drops in the X direction in turn, and the step length determination method is similar to step 1.

Step 3: Each particle moves in the direction in which the objective function descends in the Y direction in turn, and the step length determination method is similar to step 1.

Step 4: Repeat steps 1, 2, and 3 many times. When the number of repetitions is large enough or the step length is small enough, turn to step 5.

Step 5: Each powder moves along the steepest descent direction, the step length is $step_i$.

Step 6: Each particle moves along the direction in which the objective function drops in the X direction, and the step length determination method is similar to step 5.

Step 7: Each particle moves in the direction in which the objective function drops in the Y direction, and the step length determination method is similar to step 5.

Step 8: Repeat steps 6, 7, and 8 in sequence. When the step length is small enough, turn to step 9.

Step 9: Each particle randomly moves up to the top or down to the bottom.

Step 10: Repeat steps one to eight, when the step length is small enough, end the program.

Using python to program the above algorithm, the result of powder layout is shown in Fig. 3. The relative density of the powder model can be used to measure the accuracy of the powder model. The calculation method of the relative density of the powder model is shown in the Eq. (7):

$$\eta = \frac{\frac{4}{3}\sum r_k^3}{r_0^2 h_{top}} \qquad (7)$$

Where $r_k$ is radius of powder whose distance to the origin is less than $r_0$.

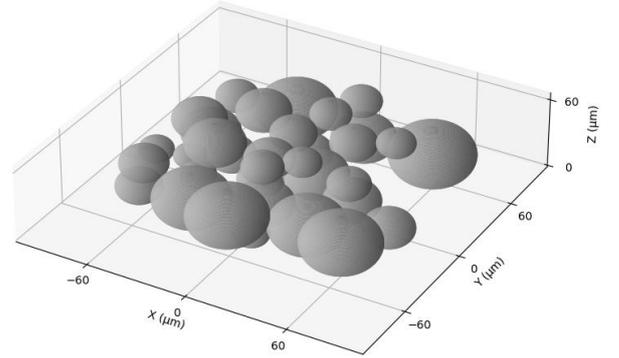

Fig. 3. Powder model.

## 2.2. The generation method of the new heat source model

During the SLM process, part of the laser beam irradiated on the powder is absorbed and the other part is reflected. The reflected energy may be absorbed by other powder particles. The entire energy absorption process is complicated. In order to explore the distribution of laser energy in the powder bed, this section discretizes the laser into several small laser beams, traces the propagation process of the laser beam, records and counts the energy absorption, finally obtains the energy distribution in the powder bed. Schematic diagram of the reflection is shown in Fig. 4.

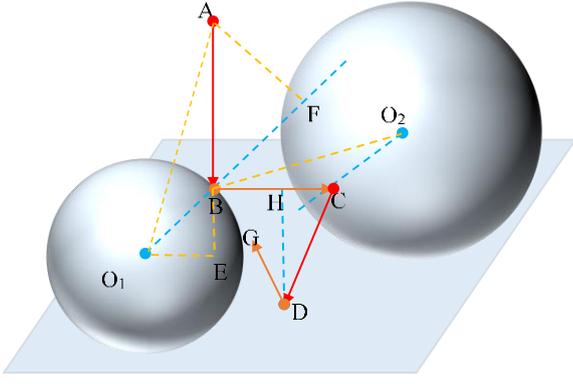

Fig. 4. Schematic diagram of the reflection path of the beam.

*2.2.1. Beam generation*

In this study, it is assumed that the laser is composed of small beams, and each small beam is represented by a straight line. The energy density distribution formula of laser [14] as follow:

$$q(x,y) = \frac{2P}{\pi r_0^2} \exp\left(\frac{-2\left((x-x_0)^2 + (y-y_0)^2\right)}{r_0^2}\right) \quad (8)$$

where $P$ is the laser power, $r_0$ is the beam waist radius of the laser, and $(x_0, y_0)$ are the coordinates of the laser center position. The power of each beam is:

$$P(x,y) = q(x,y) \cdot s \quad (9)$$

$$s = \frac{2\pi r_0}{n} \quad (10)$$

where $n$ is the number of assumed small beams, and $s$ is the cross-sectional area of beam.

*2.2.2. The reflection of the light beam on the particle*

As shown in the Fig.4, A is the initial point of the incident light, $\vec{e}_{AB}$ is the direction. B is the starting point of the beam reflection on the particle, $\vec{e}_{BC}$ is the direction. B and $\vec{e}_{BC}$ are calculated by:

$$\overrightarrow{AE} = \vec{e}_{AB} \cdot \overrightarrow{AO_1} \quad (11)$$

$$\left|\overrightarrow{O_1E}\right| = \sqrt{\left|\overrightarrow{AO_1}\right|^2 - \left|\overrightarrow{AE}\right|^2} \quad (12)$$

$$\left|\overrightarrow{BE}\right| = \sqrt{\left|\overrightarrow{BO_1}\right|^2 - \left|\overrightarrow{O_1E}\right|^2} \quad (13)$$

$$\left|\overrightarrow{AB}\right| = \left|\overrightarrow{AE}\right| - \left|\overrightarrow{BE}\right| \quad (14)$$

$$B = A + \left|\overrightarrow{AB}\right| \cdot \vec{e}_{AB} \quad (15)$$

$$\overrightarrow{BF} = \vec{e}_{O_1B} \cdot \overrightarrow{BA} \cdot \vec{e}_{O_1B} \quad (16)$$

$$\vec{e}_{BC} = \frac{\overrightarrow{AB} + 2\overrightarrow{BF}}{\left|\overrightarrow{AB} + 2\overrightarrow{BF}\right|} \quad (17)$$

*2.2.3. The reflection of the beam on the surface of the substrate*

As shown in Fig.4, C is the initial point of the incident light, $\vec{e}_{CD}$ is the direction. D is the starting point of the beam reflection on the surface of the substrate, $\vec{e}_{DG}$ is the direction. D and $\vec{e}_{DG}$ are calculated by:

$$D = C + \lambda \cdot \vec{e}_{CD} \quad (18)$$

$$\lambda = \frac{z_c}{\vec{e}_{CD} \cdot (0,0,-1)} \quad (19)$$

$$\vec{e}_{DG} = \frac{\overrightarrow{CD} + 2\overrightarrow{DH}}{\left|\overrightarrow{CD} + 2\overrightarrow{DH}\right|} \quad (20)$$

*2.2.4. Conditions for the end of beam propagation*

The remaining energy of beam propagation is determined by Eq. (21), where ε is the absorption rate, which equal to 0.3 [15]. The tracking of the current beam propagation is ended when the remaining laser energy is less than 1%, or the beam shoots out of the statistical range. Using python for programming, the propagation path of a single beam is shown in Fig. 5.

$$P_{i+1} = (1-\varepsilon)P_i \quad (21)$$

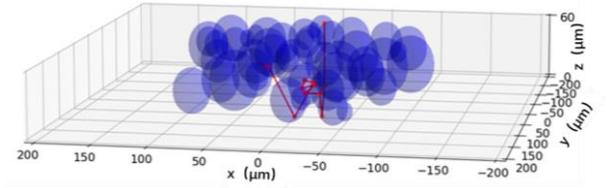

Fig. 5. Schematic diagram of a single laser beam propagating in the powder bed.

*2.2.5. Energy statistics and the change of the powder layer thickness after fusion*

The calculation steps of energy statistics are as follows:

Step 1. Divide the area into a number of hexahedral units, and calculate the energy in each unit.

Step 2. Get the average value by multiple running the program.

Step 3. Consider the change of the powder layer thickness after fusion, and adjust the energy distribution in the thickness direction. The detailed explanation as follow.

Since the density of the powder is less than the material, the actual layer thickness of the powder is different from the preset thickness except for the first layer. The schematic diagram of the thickness change of the powder layer after fusion is shown in Fig. 6.

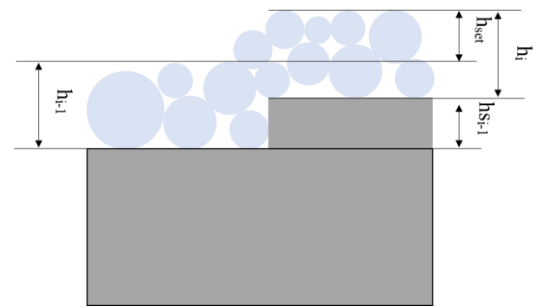

Fig. 6. Schematic diagram of the thickness change of the powder layer after fusion.

According to Fig. 6, equals are concluded as fllow:

$$h_i = h_{set} + (1-\eta) \cdot h_{i-1} \tag{22}$$

$$h_1 = h_{set} \tag{23}$$

Transform Eq. (22):

$$h_i - \frac{h_{set}}{\eta} = (1-\eta) \cdot \left(h_{i-1} - \frac{h_{est}}{\eta}\right) \tag{24}$$

$$a_i = h_i - \frac{h_{set}}{\eta} \tag{25}$$

$$a_{i-1} = h_{i-1} - \frac{h_{set}}{\eta} \tag{26}$$

$$\therefore \frac{a_i}{a_{i-1}} = 1-\eta \tag{27}$$

$$\because 0 < 1-\eta < 1 \tag{28}$$

$$\therefore \lim_{i \to \infty} a_i = 0 \tag{29}$$

$$\therefore \lim_{i \to \infty} h_i = \frac{h_{set}}{\eta} \tag{30}$$

$$\therefore \lim_{i \to \infty} hs_i = h_{set} \tag{31}$$

where $h_i$ is the height between the top surface of the powder and the upper flat of the previous layer that had been scanned by laser, $h_{set}$ is the preset powder layer thickness, and $\eta$ is the relative density.

$$\eta = \frac{\rho_p}{\rho} \tag{32}$$

where $\rho$ is the density of TC4 material, $\rho_p$ is the density of TC4 powder, $hs_i$ is the height increased of sample after fusion of the i-th layer.

In the FEM simulations, the thickness change of the powder layer cannot be calculated easily, so the layer thickness in the FEM is $h_{set}$. From the above analysis, the thickness of the powder layer used to calculate the energy distribution is $h_{set}/\eta$. In this study, $h_{set}$ is 30μm. According to section 4.1 that the $\eta$ gotten by simulates is 47.3%, and the $\eta$ gotten by experiments is 51.2%. Considering the convenience of subsequent calculations, assuming $\eta$ is 50%.

*2.2.6. New heat source model construction*

Using python to program the above method, the energy distribution after once calculation is shown in Fig. 7. As can be seen in Fig. 7 the energy distribution is not uniform, the energy is mainly concentrated on the surface of the upper hemisphere of the powder particles. The result indicates that the energy distribution strictly depends on the powder layout. In order to improve the uniformity of the energy distribution, obtaining the average results by running program multiple times. After running 1000 times, the energy distribution results remain basically constant. The results of the energy distribution in the depth direction by repeated 1000 times are shown in Fig. 8.

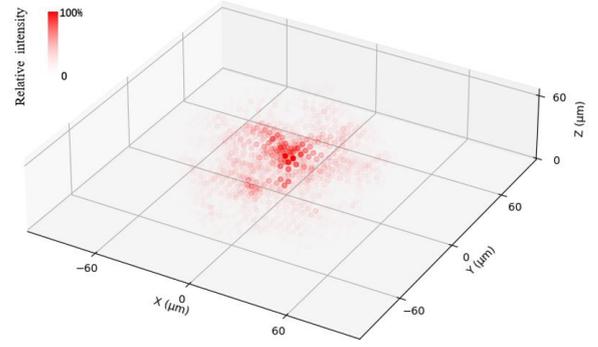

Fig. 7. Energy distribution after once calculation.

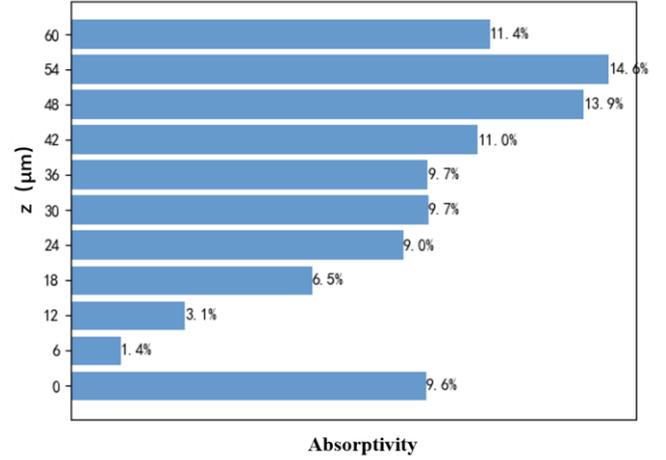

Fig. 8. Energy distribution in the depth direction (z=0 represents the surface of the substrate, z=60 represents the top of the powder layer).

The surface of the substrate absorbed the energy by the plane, but the powder absorbed the energy by the volume, the two parts of energy should be discussed separately. As can be seen from Fig. 8 that the energy absorbed by the surface of the substrate is 9.6% of the total absorbed energy, which is similar to the 11.1% reported in the literature [13], which preliminarily confirms the effectiveness of the algorithm.

The energy distribution diagram absorbed by the surface of the substrate is shown in Fig. 9:

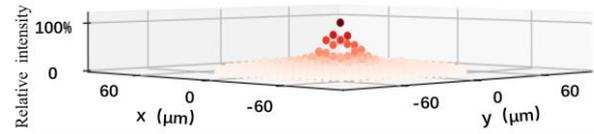

Fig. 9. Energy distribution on the surface of the substrate.

It can be seen from Fig. 9 that the energy absorbed by the surface of the substrate approximately satisfies the Gaussian distribution, so Gaussian heat source is used for surface of substrate part, and the formula [16] is:

$$q_1 = \frac{6\sqrt{3}P_1}{\pi\sqrt{\pi}r_0^2 c} \cdot e^{\left(-\frac{3(x^2+y^2)}{r_0^2} - \frac{3z^2}{c^2}\right)} \tag{33}$$

where $P_1$ is the power absorbed by the surface of the substrate.

The cloud diagram of the energy absorbed by the powders part is shown in Fig. 10.

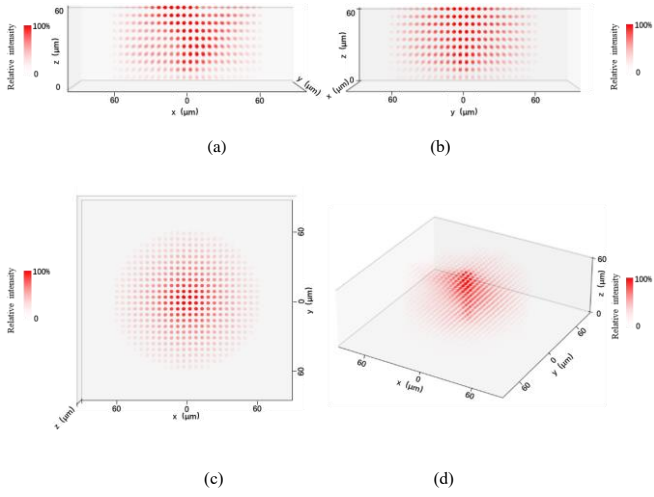

Fig. 10. Energy distributions from different perspectives after 1000 times calculations in the powder layer;(a) Front view of the energy distribution; (b) Left view of the energy distribution; (c) Top view of the energy distribution; (d) Three-dimensional view of the energy distribution.

As can be seen from Fig. 10(a) and Fig.10 (b) that the energy distribution on the vertical plane has a decreasing trend from top to bottom and from the center to both sides. As can be seen from Fig. 10(c) that the energy distribution in the horizontal direction gradually decreases from the center to the periphery. As shown in Fig. 10(d) that there is a cone-shaped high-energy distribution area in the center of the powder, which should be the reason of the formation of the keyhole during SLM processing. In order to study the distribution of energy in the horizontal direction, the energy of a layer is extracted for statistics, and the result is shown in Fig. 11:

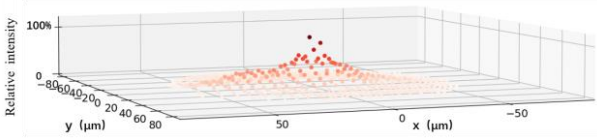

Fig. 11. Energy distribution diagram on the horizontal plane of the powder layer.

As shown in Fig. 11 that the energy absorbed by the powder approximately satisfies the Gaussian distribution in the horizontal direction. The Gaussian distribution is:

$$q = \frac{2P_2}{\pi r_0^2} \cdot e^{\left(-\frac{2(x^2+y^2)}{r_0^2}\right)} \tag{34}$$

where $P_2$ is the power absorbed by the powder part.

The analysis in the previous section shows that the thickness of the powder layer before melting is 60μm, the thickness after fusion is 30μm, and the two number are not equal. In order to match them, the Z axis of the heat source model should be changed to 50% (η). In order to obtain the relationship between the energy absorptivity and the depth in the powder part, a fifth-degree polynomial (Eq. (35)) is used for fitting, and the fitting result is shown in Fig. 12.

$$f(z) = az^5 + bz^4 + cz^3 + dz^2 + ez + f \tag{35}$$

where $a, b, c, d, e, f$ are undetermined coefficients, which are calculated by the ordinary least squares.

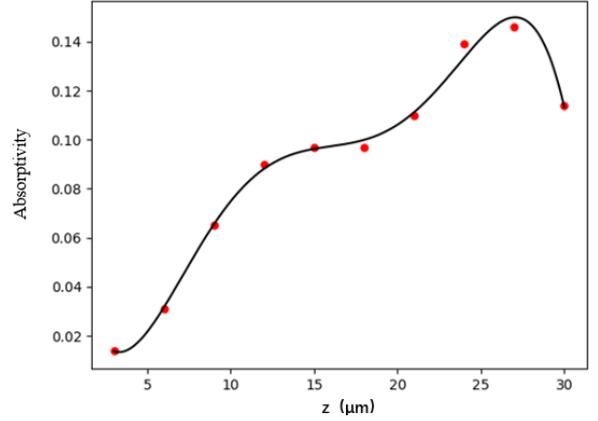

Fig. 12. Fifth-degree polynomial fitting curve of the relationship between the energy absorptivity and the depth.

As can be found from Fig. 12 that the absorptivity by the powder part first increases and then decreases in the depth direction. The trend is similar to the results reported in the literature [13], which proves the effectiveness of the algorithm. In addition, the absorptivity has two extremums at z=15 and z=27. This is due to the fact that energy is mostly distributed in the upper hemisphere of the powder particles (as can be seen from Fig. 7). The powders are usually staggered distribution in two layers in the height direction by the powder particles radius and the thickness, therefore two extremums phenomenon is reasonable. The absorptivity of the top layer is slightly lower, it is also reasonable because the upper part of the spherical particles in the top layer cannot absorb the reflected laser energy.

According to the above analysis, a heat source model that satisfies the Gaussian distribution in the horizontal direction and changes in the vertical direction according to the law of the function is used to characterize the distribution of the absorbed energy of the powders part. The equation can be written as:

$$q_2 = \frac{2kP_2}{\pi r_0^2 h_{set}} \cdot e^{\left(-\frac{2(x^2+y^2)}{r_0^2}\right)} \cdot f(z) \tag{36}$$

$$P_2 = \iiint q_2 dv \tag{37}$$

where $k$ is the undetermined coefficient. The value of $k$ can be calculated by Eq. (36) and Eq. (37).

The final heat source model is shown in Eq. (38), which consist of the surface of substrate part heat source model and powder part heat source model.

$$q=\begin{cases} \dfrac{6\sqrt{3}P_1}{\pi\sqrt{\pi}r_0^2 c}\cdot e^{\left(-\dfrac{3\left((x-x_0-vt)^2+(y-y_0)^2\right)}{r_0^2}-\dfrac{3(z-z_0)^2}{c^2}\right)}, z\leq z_0 \\ \dfrac{2kP_2}{\pi r_0^2 h_{set}}\cdot e^{\left(-\dfrac{2\left((x-x_0-vt)^2+(y-y_0)^2\right)}{r_0^2}\right)}\cdot\left(a(z-z_0)^5+b(z-z_0)^4+c(z-z_0)^3+d(z-z_0)^2+e(z-z_0)+f\right), z>z0 \end{cases} \quad (38)$$

In summary, a new heat source model is proposed, which consists of two parts, the heat source on the surface of the substrate and the heat source on the powder. The energy of the surface of the substrate accounts for 9.6%, which is characterized by a Gaussian heat source model. The energy of the powder part accounts for 90.4%, which is characterized by the heat source model that the horizontal direction satisfies the Gaussian distribution, and the vertical direction changes according to the function law.

### 2.3. Finite element temperature field model of SLM

#### 2.3.1. Modeling process

Use ABAQUS to build finite element model. The size of the substate is 1*0.5*0.27mm, and the size of the powder layer is 1*0.5*0.03mm. In order to improve the accuracy of simulation calculation and reduce the amount of calculation, the powder layer is meshed into a uniform regular hexahedral grid, and the substrate is meshed by single bias method. The farther away from the processing area, the sparser element density. The DFLUX subroutine is used to load the heat source model that proposed in this paper. The process is shown in Fig. 13.

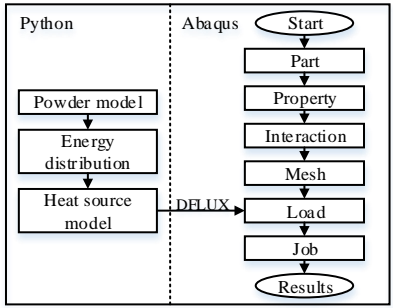

Fig. 13. Modeling process

#### 2.3.2. Material parameters

The material used in this study is TEKNA's TC4 powder, and the chemical composition is shown in Table 1. The powder is sieved by a 300-mesh sieve under the protection of nitrogen using a powder sieving machine from Dahan Machinery Company. Malvern Masterizer2000 was used to detect the powder size distribution. The powder size distribution is shown in Fig. 14.

**Table 1**

Chemical Composition of TC4 powder.

| | Chemical Composition | | | | | | | |
|---|---|---|---|---|---|---|---|---|
| Element | Al | V | Fe | Y | C | O | N | H | Ti |
| Result (wt%) | 6.41 | 4.02 | 0.14 | 0.001 | 0.007 | 0.075 | 0.016 | 0.003 | Balance |

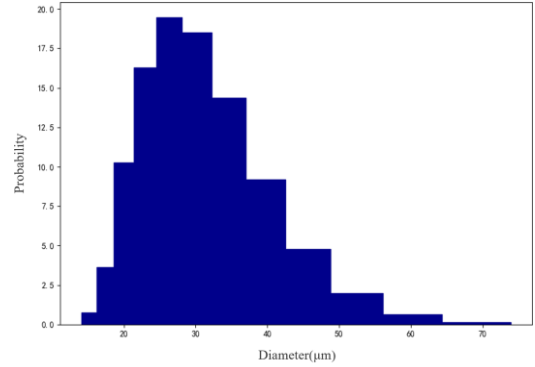

Fig. 14. Powder size distribution.

The density $\rho$ of TC4 material is 4450 $kg/m3$, and the thermal conductivity and heat capacity varying with temperature are shown in Fig. 15.

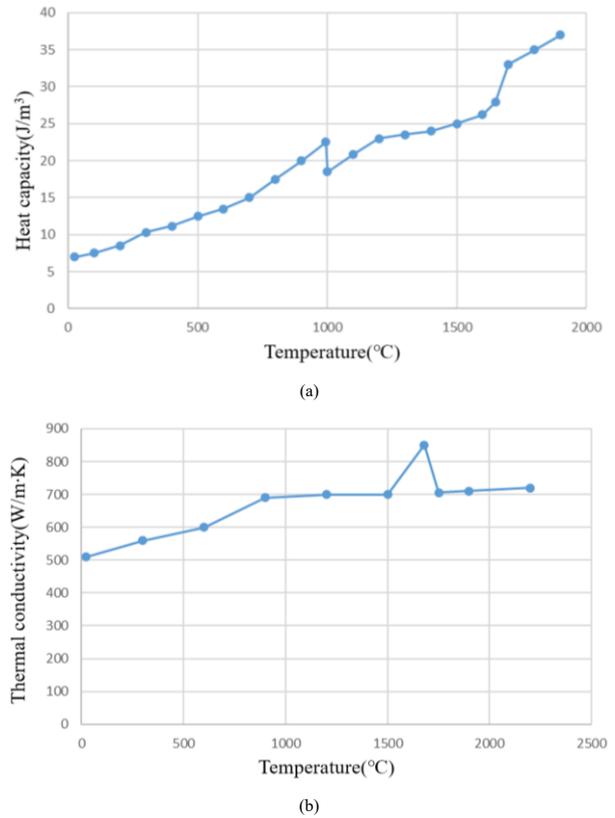

Fig.14. Thermal material properties of TC4; (a)Thermal conductivity[17]; (b) Heat capacity [18].

#### 2.3.3. Initial state and boundary conditions

The initial temperature of the powder layer, the substrate and the environment are all 118°C.

Considering the influence of convection on the upper surface of the molten pool, the convection coefficient is taken as 24W/(m²K)[19].Consider the influence of the upper surface radiation of the molten pool on the temperature field.

Ignore the influence of Marangoni convection and vaporization recoil pressure.

## 3. Experiments

### 3.1. Measuring the relative density of the powder layer

In order to measure the relative density of the powder layer, a powder scattering experiment was carried out with the tool as shown in Fig. 16. The tool consists of three feeler gauges with a thickness of 60μm. No. 1 and No. 2 feeler gauges as a guide for the scraper were stacked on No. 3. to ensure that the thickness of the powder layer is 60μm. The area between No. 1 and No. 2 feeler gauges is the powder scattering area, and the volume $V$ is 82.8mm³. Use a scraper to scattering the powder particles at a speed of 200 mm/s. The mass of the tool is $m_1$, $m_1$=2.8875g, the mass after powder coating is $m_2$, $(m_2-m_1)$ is the mass of the powder layer. The calculation formula of relative density is Eq. (39):

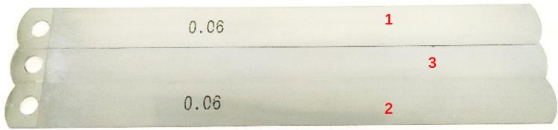

Fig. 16. The tool for scattering power layer.

$$\eta = \frac{m_2 - m_1}{V \cdot \rho} \tag{39}$$

### 3.2. Sample preparation

In order to verify the accuracy of the temperature field simulation model, it is necessary to make a single-channel molten pool observation sample. A SLM metal 3D printer model EP-M250 of Easy Plus 3D was used for samples preparation. The samples are prepared in two steps.

Step 1: Using 180W, 1000mm/s process parameters, print 5*5*6mm cuboids on the substrate, a schematic diagram of the first step is shown in Fig. 17(a).

Step 2: Scattering a thickness is 60μm powder layer on the printed cuboids, then laser scan the single-channel perpendicular to the scan direction of the first step. The schematic diagram of the second step is shown in Fig. 17(b). The scanning process parameters are shown in Table 2. The actual samples are shown in Fig. 18.

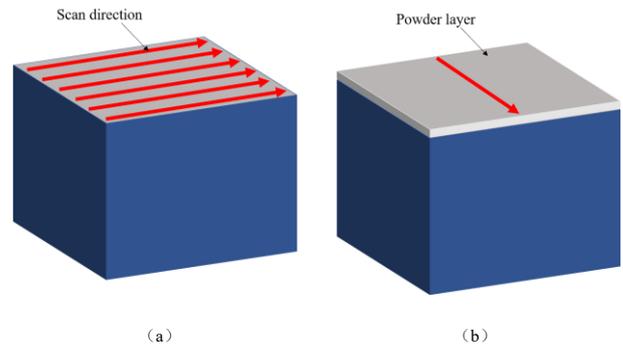

(a)　　　　　　　　　　　(b)

Fig. 17. Schematic diagram of sample preparation (a) Step 1; (b) Step 2.

**Table 2**

Parameters used in experiments.

| NO. | Power(w) | velocity(mm/s) |
|---|---|---|
| 1 | 160 | 1000 |
| 2 | 170 | 1000 |
| 3 | 180 | 1000 |
| 4 | 190 | 1000 |
| 5 | 200 | 1000 |
| 6 | 180 | 800 |
| 7 | 180 | 900 |
| 8 | 180 | 1000 |
| 9 | 180 | 1100 |
| 10 | 180 | 1200 |
| 9 | 180 | 1100 |
| 10 | 180 | 1200 |

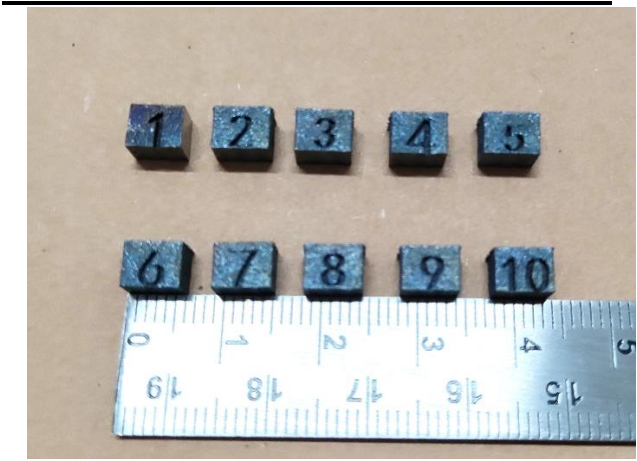

Fig. 18. The photo of the samples.

### 3.3. Sample processing

Sample processing: Use 200, 400, 800, 1600, 3000, 5000, 7000 mesh sandpaper to polish the cross-sections perpendicular to the scanning direction of the samples. Observe the cross-section with the model SMZ1270 optical microscope.

## 4. Results and discussion

### 4.1. Powder model verification

The comparison between the TC4 powder models and the results of the scattering experiments is shown in Fig. 19:

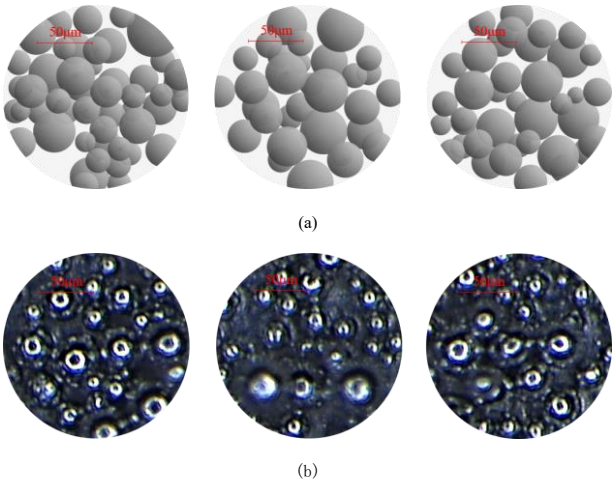

(a)

(b)

Fig. 19. Comparison between the powder models and the experiments; (a) Powder models; (b) Photos observed by microscope.

As can be seen from Fig. 19 that the powder models are similar to the experimental results.

To further prove the validity of the model, the average relative density of the powder model is 47.3%. And the experimental results of the relative density of the powder layer are shown in Table 3. As can be seen that the average relative density of the powder layer obtained in the experiments is about 51.2%. Compare the results of experiments and model, there is a 3.9% error between them. There may be two reasons for the error: The first reason is the error between the smooth plane assumed in the model and the surface of the experimental tool; Another one may be the local optima produced by the optimization algorithm. Since the error is only 3.9%, the powder model is considered valid.

**Table 3**

Results of the relative density of the powder layer by experiments.

|  | Test1 | Test2 | Test3 | Average |
|---|---|---|---|---|
| $m_2$(g) | 3.0627 | 3.0912 | 3.0746 | 3.0762 |
| $\eta$ | 47.5% | 55.3% | 50.1% | 51.2% |

### 4.2. Results and analysis

Use the model proposed in this paper and the parameters in Table 2 to simulate the temperature fields. The simulations and experimental results are shown in Fig. 20 and Fig. 21. The experimental and simulation results of the molten pool width are shown in Table 4, the experimental and simulation results of the connect width are shown in Table 5.

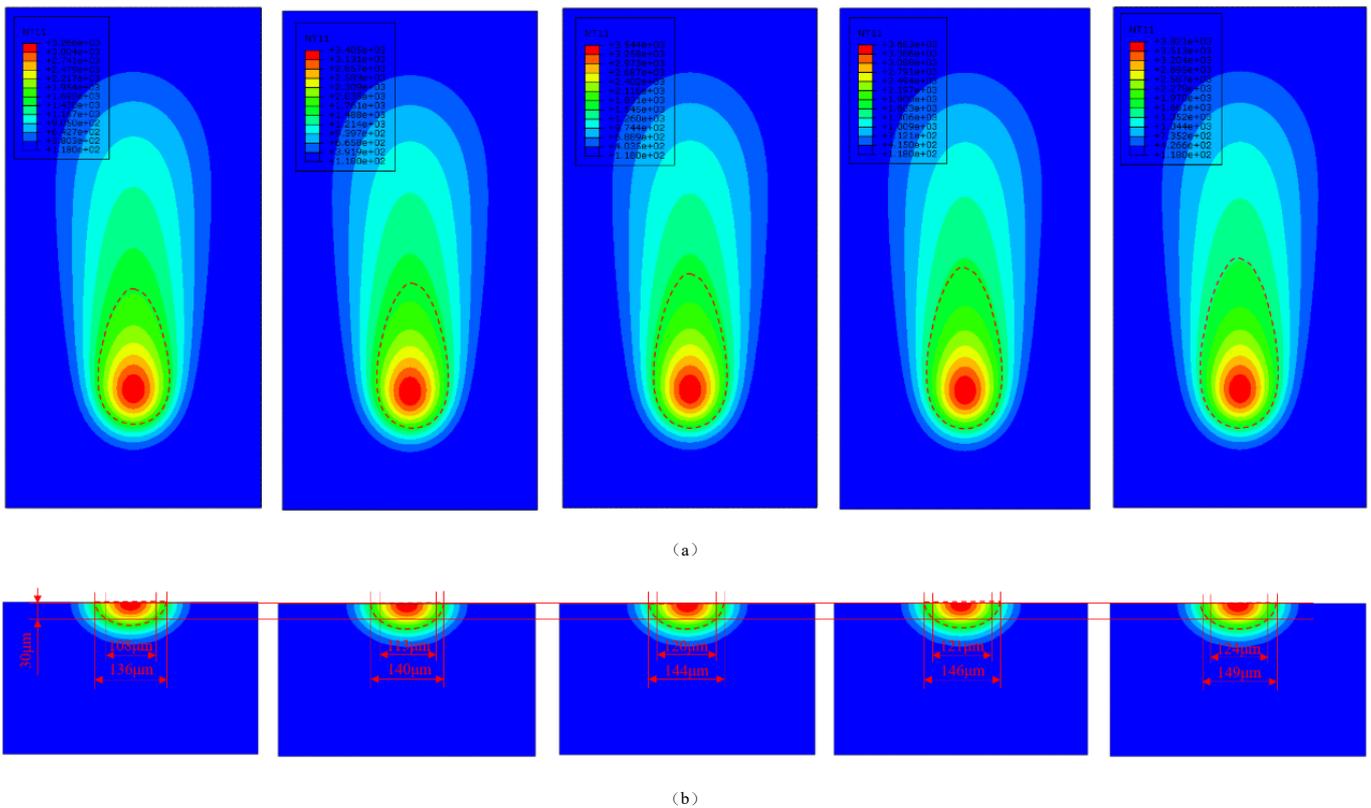

(a)

(b)

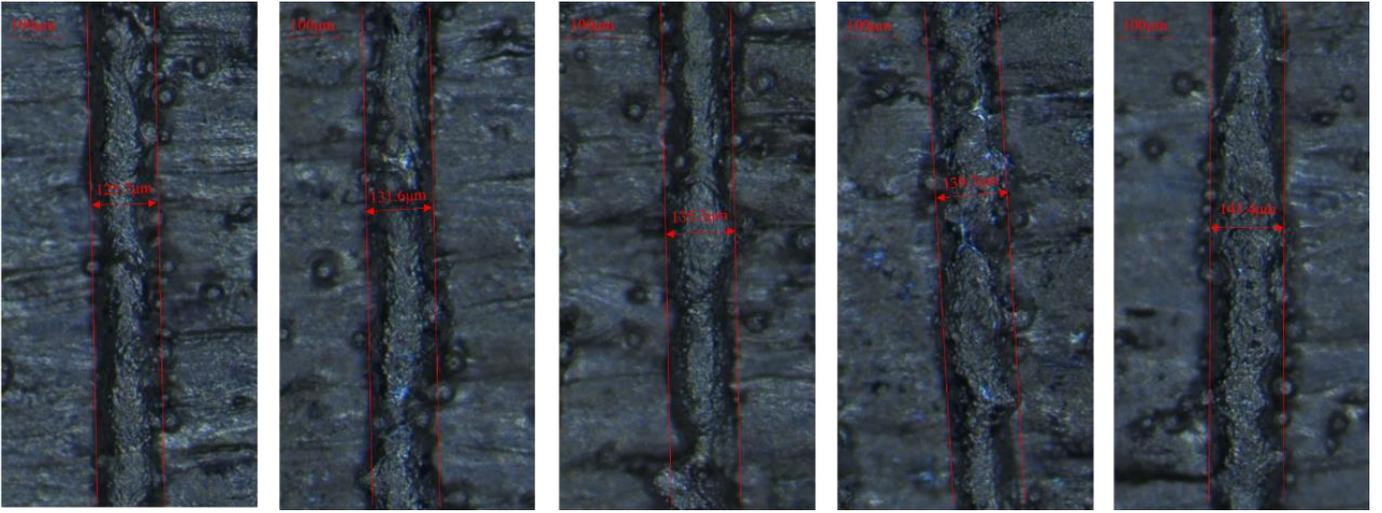

(c)

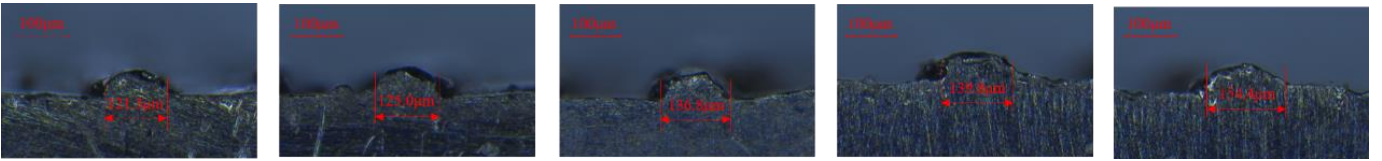

(d)

Fig. 20. Simulation and experimental results for scanning speeds of 1000 mm/s and laser power of 160, 170, 180, 190 and 200W; (a) simulated temperature distribution on surface of powder layer; (b) simulated cross-section of melt pool; (c) experimental top-view images of melt tracks; (d)experimental cross-sectional images of melt tracks.

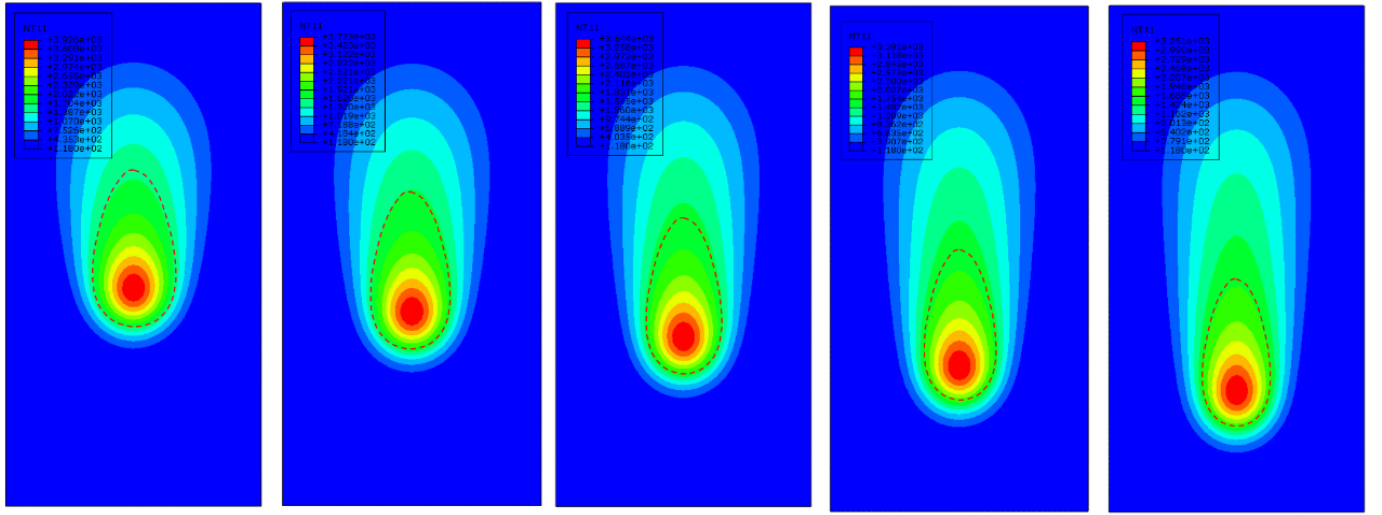

(a)

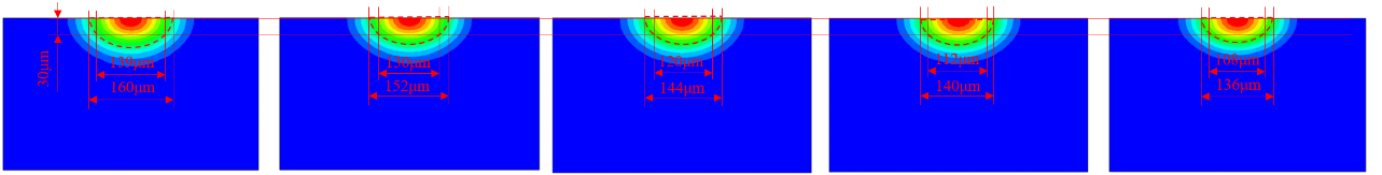

(b)

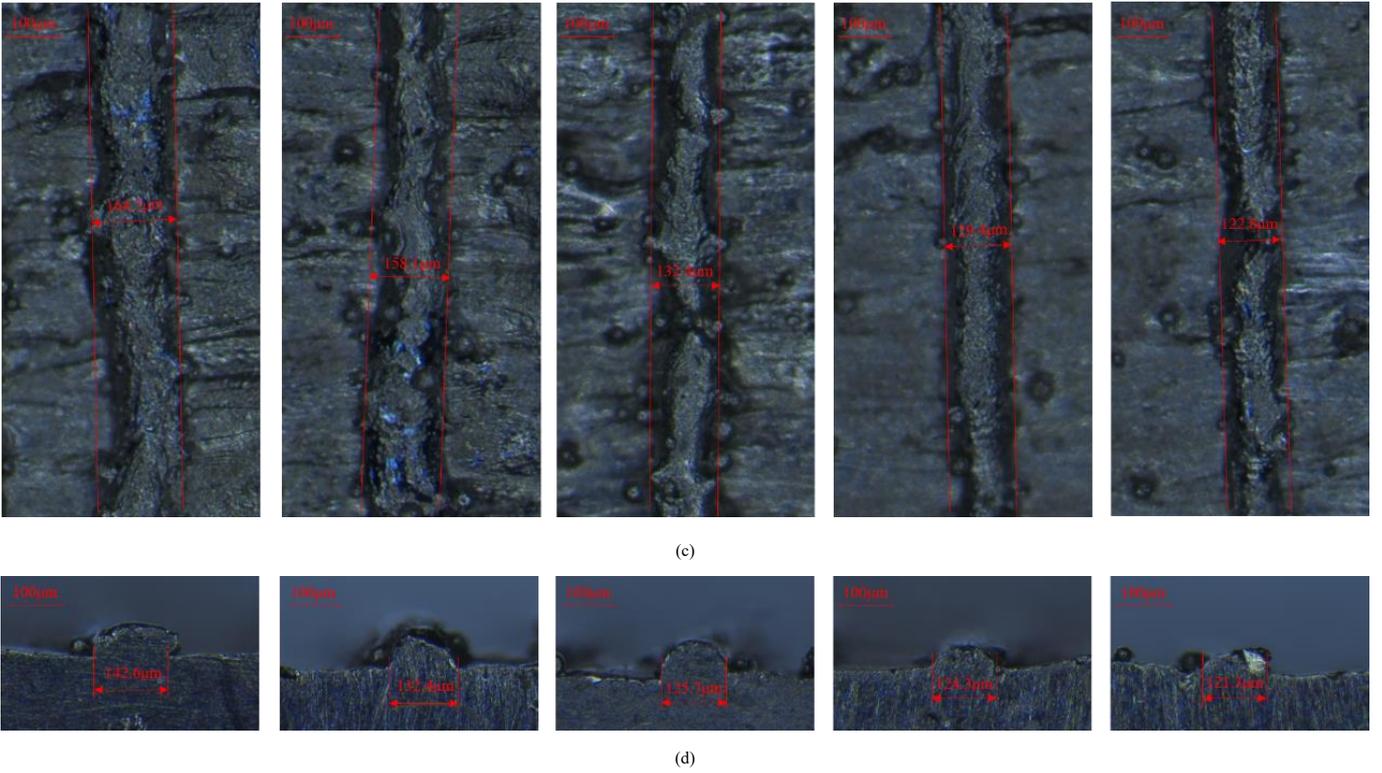

(c)

(d)

Fig. 21. Simulation and experimental results for laser power of 180 W and scanning speeds of 800, 900,1000,1100 and 1200 mm/s; (a) simulated temperature distribution on surface of powder layer; (b) simulated cross-section of melt pool; (c) experimental top-view images of melt tracks; (d)experimental cross-sectional images of melt tracks.

**Table 4**

Comparison of the experimental results for melt pool width with the simulation results.

|  | 1 | 2 | 3 | 4 | 5 | 6 | 7 | 8 | 9 | 10 |
|---|---|---|---|---|---|---|---|---|---|---|
| Experimental results(μm) | 125.7 | 131.6 | 135.3 | 139.7 | 143.4 | 164.7 | 158.1 | 132.4 | 129.4 | 122.8 |
| Simulation results(μm) | 136 | 140 | 144 | 146 | 149 | 160 | 152 | 144 | 140 | 136 |
| Average relative error |  |  |  |  | 6.4% |  |  |  |  |  |

**Table 5**

Comparison of the experimental results for melt pool connect width with the simulation results.

|  | 1 | 2 | 3 | 4 | 5 | 6 | 7 | 8 | 9 | 10 |
|---|---|---|---|---|---|---|---|---|---|---|
| Experimental results(μm) | 120 | 115 | 124 | 145 | 131 | 141 | 131 | 127 | 117 | 126 |
| Simulation results(μm) | 108 | 113 | 120 | 121 | 124 | 139 | 130 | 120 | 112 | 108 |
| Average relative error |  |  |  |  | 9.6% |  |  |  |  |  |

As can be seen from Fig. 20(a), Fig. 20(b), Table 4 and Table 5 that when other parameters are the same, as the power increases from 160w to 200w, the maximum temperature of the molten pool gradually rises from 3266°C to 3821°C. The width increased from 136μm to 149μm, and the connect width increased from 108μm to 124μm. The maximum temperature of the molten pool obtained by the simulation is higher than the vaporization temperature of the TC4 powder [20], and splashing of the molten liquid may occur. Since the flow and the evaporation are not considered, the maximum temperature of this model may be higher than the actual temperature [21].

As shown in Fig. 21 (a), Fig. 21(b) and Table 4 and Table 5, when other parameters are the same, as the speed increases from 800mm/s to 1200mm/s, the maximum temperature of the molten pool decreases from 3926 ℃ to 3251 ℃, the width of the molten pool is reduced from 160μm to 136μm, and the connect width is reduced from 139μm to 108μm.

As can be seen from Fig. 20(c) and Fig. 21(c) that there are many powder particles embedded in the melt pool on both sides. A small amount of particles are found on the top of the melt pool, which may be caused by the flow of the molten pool. In addition, As can be found that the morphology of the melt pool is more uniform in the areas where the surface of the substrate is flatter, while the morphology of the molten pool changes greatly in the areas where the surface of substrate is rougher. The reason is when the surface roughness of the substrate is bigger, the drop between the peaks and valleys is larger, resulting in a bigger difference in the amount of powder distributed on the peaks and valleys, then the volume of the melt formed by the laser irradiation at the peaks and valleys are more different, and the height difference between the peaks and valleys promotes the molten liquid to flow to the valleys, which ultimately results in poor uniformity of the newly formed molten pool.

As shown in Fig. 21(d) and Fig. 21(d) the connect between the melt pool and the substrate is mostly a "slope" connection, which may be caused by the flow of the melt pool. This reason also caused the connect width measured by the experiments to be wider than the predicted by the simulations.

Comparison of this study with Gusarov[12] and Tran[13] is shown in Table 6.

**Table 6**

Comparison of this study for prediction errors with previous.

|  | Guasarov | Tran | present |
|---|---|---|---|
| Material | 316L | 316L | TC4 |
| Radius of laser beam, r0(μm) | 50μm | 50μm | 60μm |
| Thickness of powder layer(μm) | 30μm | 30μm | 30μm |
| Average relative error of pool width | 4.6% | 6.6% | 6.4% |
| Average relative error of connect width | 17.8% | 14.0% | 9.6% |

As can be seen from Table 6 that the average error of the pool width between the simulation results and the experimental results in this study is 6.4%, and the average error of the connect width is 9.6%. The average error of Gusarov's pool width is 4.6%, the average error of connect width is 17.8%, the average error of Tran's pool width is 6.6%, and the average error of connect width is 14%. It can be concluded that the model proposed in this study is better than Gusarov and Tran in predicting the connect width. It may be that the heat source model proposed in this study has considered in detail the distribution of energy at the connection on the powder layer and the substrate, so the prediction at the connection has a higher accuracy.

## 5. Conclusion

This paper proposed a finite element modeling method for predicting the shape of the SLM melt pool. Notably, the simulation used a new heat source model, which takes into account the distribution of laser energy between the surface of the substrate and the powder layer. In particular, an optimization method was proposed to simulate powder layout, which could avoid the error caused by estimated powder physical parameters. The effectiveness of the powder model was verified by related experiments.

The thickness change of the powder layer after fusion was demonstrated. When the layer thickness is set as $h_{set}$, the actual layer thickness of the powder layer should be $h_{set}/\eta$.

Python programming was used to count the distribution of energy on the surface of substrate and powder layer. According to the distribution of energy, the heat source model is consisted of two parts, one is a Gaussian heat source model for thr surface of the substrate, the other is a heat source model with Gaussian distribution in the horizontal direction and the vertical direction changing according to the law of the function for the powder layer.

The proposed model was used to predict the geometry of the SLM melt pool, and the results show that the prediction results are consistent with the experimental results. However, the flow and evaporation of the melt pool were not considered, the maximum temperature of the melt pool in the simulation results is higher.

## Acknowledgements

Supported by Sichuan Science and Technology Program (grant number: 2020YFG0407).